\documentclass[aps,prx,preprint]{revtex4-1}

\usepackage{amsmath, amsthm, amssymb, graphicx}
\usepackage{color}
\usepackage{multirow}
\usepackage[normalem]{ulem}
\usepackage{float}
\usepackage{bm}
\usepackage{siunitx}
\usepackage{booktabs}
\usepackage{array}
\usepackage[utf8x]{inputenc}

\begin{document}

\def\be{\begin{equation}}
\def\ee{\end{equation}}
\def\bf{\begin{figure}}
\def\ef{\end{figure}}
\def\bea{\begin{eqnarray}}
\def\eea{\end{eqnarray}}
\def\bml{\begin{mathletters}}
\def\eml{\end{mathletters}}
\def\b{\bullet}
\def\eqn#1{(~\ref{eq:#1}~)}
\def\no{\nonumber}
\def\av#1{{\langle  #1 \rangle}}
\def\m{{\rm{min}}}
\def\M{{\rm{max}}}

\title{Feedback control of surface roughness in a one-dimensional KPZ growth process}

\author{Priyanka}%
 \email{pri2oct@vt.edu}
\affiliation{Department of Physics and Center for Soft Matter and Biological Physics, Virginia Tech, Blacksburg, VA 24061-0435, USA}
\author{Uwe C. T{\"a}uber}%
 \email{tauber@vt.edu}
\affiliation{Department of Physics and Center for Soft Matter and Biological Physics, Virginia Tech, Blacksburg, VA 24061-0435, USA}
\author{Michel Pleimling}
 \email{pleim@vt.edu}
\affiliation{Department of Physics and Center for Soft Matter and Biological Physics, Virginia Tech, Blacksburg, VA 24061-0435, USA}
\affiliation{Academy of Integrated Science, Virginia Tech, Blacksburg, VA 24061-0563, USA}
 
\date{\today}
%


\date{\today}

\begin{abstract}
Control of generically scale-invariant systems, i.e., targeting specific cooperative features in non-linear stochastic interacting systems with many degrees of freedom subject to strong fluctuations and correlations that are characterized by power laws, remains an important open problem.
We study the control of surface roughness during a growth process described by the Kardar--Parisi--Zhang (KPZ) equation 
in $(1+1)$ dimensions. 
We achieve the saturation of the mean surface roughness to a prescribed value using non-linear feedback control. 
Numerical integration is performed by means of the pseudospectral method, and the results are used to investigate the coupling effects of controlled (linear) and uncontrolled (non-linear) KPZ dynamics during the control process. 
While the intermediate time kinetics is governed by KPZ scaling, at later times a linear regime prevails, namely the relaxation towards the desired surface roughness. 
The temporal crossover region between these two distinct regimes displays intriguing scaling behavior that is characterized by non-trivial exponents and involves the number of controlled Fourier modes. 
Due to the control, the height probability distribution becomes negatively skewed,  which affects the value of the saturation width.
\end{abstract}



\maketitle

\section{Introduction}{\label{Sec 1}}

Non-linear stochastic processes are abundant in fields as diverse as physics, economics, the life sciences, and engineering. 
In physics, some prominent examples encompass Brownian particles, kinetic roughening of surfaces, and reaction-diffusion systems, to name but a few, yet all of the above find wide and crucial applications in science and technology.
Generic scale invariance features prominently far away from thermal equilibrium: 
In such dynamical non-linear stochastic interacting systems with many degrees of freedom, one frequently observes strong fluctuations and correlations that are characterized by power laws. 
Control of generically scale-invariant systems, i.e., targeting for desired specific macroscopic properties, remains an important open problem.

Non-equilibrium growth processes of interfaces and kinetic roughening of surfaces are of technological relevance in areas ranging from thin films and nanostructures to biofilms and diffusion fronts. 
In many of these processes the fluctuating and roughening interface can be described on a mesoscopic level by stochastic non-linear partial differential equations, as for example the celebrated Kardar--Parisi--Zhang (KPZ) \cite{Kardar86} or  Kuramoto--Sivashinsky \cite{Kuramoto78,Sivashinsky80} equations.
Because of their many areas of applications and technological importance, various approaches have been discussed in the literature to control these growth processes and achieve a pre-determined outcome as for example an a priori determined interface width \cite{Armaou00,Lou03,Lou06,Gomes15,Gomes17}. 
The simplest schemes essentially suppress the non-linearity, which results in the efficient control of a linear stochastic differential equation. 
More subtle approaches aim at retaining some of the non-linear features, see for example Ref.~\cite{Gomes17}, but even these schemes ultimately lose the characteristic defining scaling properties of the underlying stochastic non-linear growth process.

In this paper we present a control protocol that in contrast both maintains distinguishing scaling features of the uncontrolled non-linear stochastic process and enforces a desired width for the kinetically roughening surface. 
Employing a pseudospectral scheme with mixed implicit-explicit time integration, we show that the coupling between controlled and uncontrolled Fourier modes yields a mix between non-linear and linear dynamics that results in remarkably non-trivial crossover scaling properties during the roughening process. 
We thus demonstrate that it is possible to precisely target a desired value for the mean interface width for a growing, roughening surface while preserving its rich and non-trivial scale-invariant dynamics prior to saturation.

Our manuscript is organized in the following way. 
Using the one-dimensional KPZ equation as our test case, we first discuss in Section~\ref{Sec 2} the pseudospectral scheme with mixed time integration, before explaining in Section~\ref{Sec 3} our non-linear feedback control protocol with periodic control actuators that we use in order to achieve a steady state with a desired surface width. 
In Section~\ref{Sec 4} we present our numerical results and show that this control scheme yields a mix of non-linear and linear dynamics that results in intriguing scaling properties characterized by non-trivial scaling exponents. 
We conclude and provide our outlook in Section~\ref{Sec 5}.

\section{Pseudospectral scheme with implicit-explicit time integration for the KPZ equation}{\label{Sec 2}}

In terms of the time-dependent local surface height function $h(x,t)$, the one-dimensional KPZ equation \cite{Kardar86} is given by
\begin{equation} \label{KPZ_eq}
\frac{\partial h}{\partial{t}} = \nu \frac{\partial^2{h}}{\partial{x^2}} + \frac{\lambda}{2} \left| \frac{\partial{h}}{\partial{x}} \right|^2 + \xi(x,t)~,
\end{equation}
where $\nu$ is the diffusion constant, $\lambda$ is the strength of the non-linearity, and $\xi(x,t)$ represents delta-correlated white noise of strength $\sigma$, i.e., $\langle\xi(x',t')\xi(x,t)\rangle=\sigma\delta(x-x')\delta(t-t')$. 
On the right-hand side of Eq.~(\ref{KPZ_eq}), the first term smoothens the growing interface, whereas the noise term tends to roughen it. 
The non-linear term describes growth enhancement induced by local height gradients.
For the study reported below we consider finite systems of size $L$, for which we employ periodic boundary conditions, $h(x+L,t)= h(x,t)$.

The quantity of interest for our work is the mean surface width or interface roughness
\begin{equation}
W(L,t) = \sqrt{\frac{1}{L} \int_{-L/2}^{L/2} \Big[ h(x,t) - \langle h(L,t) \rangle \Big]^2 dx} \ ,
\label{Width_eq}
\end{equation}
where $\langle h (L,t) \rangle = \int_{-L/2}^{L/2} h(x,t) dx / L$ denotes the mean height at time $t$.
In a finite system this quantity displays for many growth processes the Family--Vicsek scaling behavior \cite{Family85,Chou09}
\begin{equation} \label{fvscaling}
W(L,t)=L^{\alpha} {\cal W}(t/L^{z})~,
\end{equation}
with the dynamical exponent $z$ and the roughness exponent $\alpha$, whereas the scaling function has the limits
${\cal W}(y) \longrightarrow 1$ for $y \longrightarrow \infty$ and ${\cal W}(y) \sim y^\beta$ for $y \ll 1$. 
The exponent $\beta$ is called growth exponent and is given by $\beta = \alpha / z$.
It follows that the time until saturation scales as $L^z$, and that the saturation width is proportional to $L^\alpha$.
For the KPZ equation in one space dimension one has $\alpha = 1/2$ and $z = 3/2$, thus verifying the general Galilean (tilt) invariance scaling relation $\alpha + z =2$ \cite{Forster77}. 
The ratio of these values yields $\beta = 1/3$. 

One characteristic of the one-dimensional KPZ equation is that its stationary height fluctuations do not depend on the non-linearity in Eq.~(\ref{KPZ_eq}) \cite{Forster77,Huse85,Krug92}, which allows to obtain exact expressions for stationary quantities. 
With the discrete Fourier modes $\hat{h}(k,t)$, $k = 2 \pi m /L$ and $m = -L/2, \cdots, L/2$, one computes the variance
\begin{equation}
\lim_{t\rightarrow\infty} \langle |\hat{h}(k,t)|^2 \rangle = \frac{\sigma}{2\nu L k^2}
\end{equation}
and the stationary width \cite{Krug92}
\begin{equation}
 \lim_{t\rightarrow\infty} W(L,t) = \sqrt{\frac{\sigma}{24 \nu}} \, L^{1/2}~.
\label{width}
\end{equation}

It will be useful in the following to work with dimensionless variables in terms of which the KPZ equation becomes
\begin{equation}
\frac{\partial h}{\partial{t}}  =\frac{\partial^2{h}}{\partial{x^2}} + \frac{g}{2 }\left| \frac{\partial{h}}{\partial{x}} \right|^2 + \eta(x,t)~,
\label{SKPZ}
\end{equation}
with the non-linear coupling $g=\lambda\sqrt{\sigma/\nu^3}$ and the delta-correlated Gaussian noise $\eta(x,t)$ with zero mean and unit variance.
Various discretization schemes have been used for the numerical investigation of the KPZ equation \cite{Amar90,Beccaria94,Lam98a,Buceta05}, and possible issues with some of these schemes are well-documented \cite{Lam98b}.
In (discrete) Fourier space, Eq.~(\ref{SKPZ}) reads
\begin{equation}
\frac{\partial {\hat{h}}(k,t)}{\partial t} = - k^2 {\hat{h}}(k,t) + Q(k,{\hat{h}}) + \hat{\eta}(k,t)~,
\label{F_KPZ}
\end{equation}
where $Q(k,{\hat{h}})$ is the non-linear term of the KPZ equation given as a convolution sum,
\begin{equation}
Q(k,{\hat{h}}) = \frac{g}{2} \sum_{q,q'} q\,q' {\hat{h}}(q,t) {\hat{h}}(q',t) \,\delta_{q+q',k}~.
\label{NLKPZ}
\end{equation}

Pseudospectral methods with explicit time integration have been used successfully in the past for the Fourier space KPZ equation (\ref{SKPZ}) \cite{Giada02,Gallego07}, and have been shown to be superior to standard real-space discretization schemes. 
The stability of the numerical integration can be further enhanced by replacing the explicit scheme with a mixed implicit-explicit method \cite{Ascher95}.
Mixed methods of time integration are widely used in fluid dynamics, for example to solve Navier-Stokes equations~\cite{Canuto88,Spalart91}, but they do not seem to have found broad application yet in surface growth processes.
In these approaches, an implicit scheme is used for the linear terms, whereas an explicit scheme is used for the non-linear contributions. 
This mixed treatment enhances the stability of the numerical integration, permitting larger time integration steps. 
We follow the scheme described in Ref.~\cite{Spalart91}, where the non-linear and noise terms are treated with a low-storage third-order Runge--Kutta scheme, whereas the standard Crank--Nicolson scheme is used for the linear diffusion term.
We also employ the $3N/2$ aliasing rule \cite{Giada02}.

In our implementation the step from $\hat{h}(t)$ to $\hat{h}(t+\delta t)$ (for these equations we do not explicitly write the dependence of $\hat{h}$ and $Q$ on $k$) is done in three sub-steps:
\begin{align} 
\hat{h}_{1} = \hat{h}(t) + \delta t \Big[ &-k^2 (\alpha_1\hat{h}(t)+\beta_1 \hat{h}_{1}) +\gamma_1 Q(\hat{h}) +\gamma_1\hat{\eta} \Big] \no\\
\hat{h}_{2} = \hat{h}_1 + \delta t &\Big[ -k^2 (\alpha_2\hat{h}_1+\beta_2 \hat{h}_{2})+\gamma_2 Q_1 \no\\
&+\zeta_1 Q(\hat{h}) +(\gamma_2+\zeta_1)\hat{\eta} \Big] \no\\
\hat{h}(t+ \delta t) = \hat{h}_2 + \delta t &\Big[ -k^2 (\alpha_3\hat{h}_2+\beta_3 \hat{h}(t+\delta t)) +\gamma_3 Q_2 \no\\
&+\zeta_2 Q(\hat{h}) +(\gamma_3+\zeta_2)\hat{\eta} \Big]~,
\end{align}
where $Q_i = Q(\hat{h}_i)$ is the non-linear term calculated at sub-step $i$. 
The coefficients must satisfy the relation 
\begin{equation}
\gamma_1 +\gamma_2+\gamma_3+\zeta_1+\zeta_2=1~.
\end{equation}
Following Ref.~\cite{Spalart91}, see the Appendix in that publication, the coefficients take on the values
\begin{eqnarray*}
&&\alpha_1=29/96;\ \alpha_2=-3/40;\  \alpha_3=1/6;\\
&&\beta_1=37/160;\ \beta_2=5/24;\ \beta_3=1/6;\\
&&\gamma_1=8/15;\ \gamma_2=5/12;\ \gamma_3=3/4;\\
&&\zeta_1=-17/60;\ \zeta_2=-5/12.
\end{eqnarray*}

\begin{figure}
 \centering \includegraphics[width=0.6\columnwidth,clip=true]{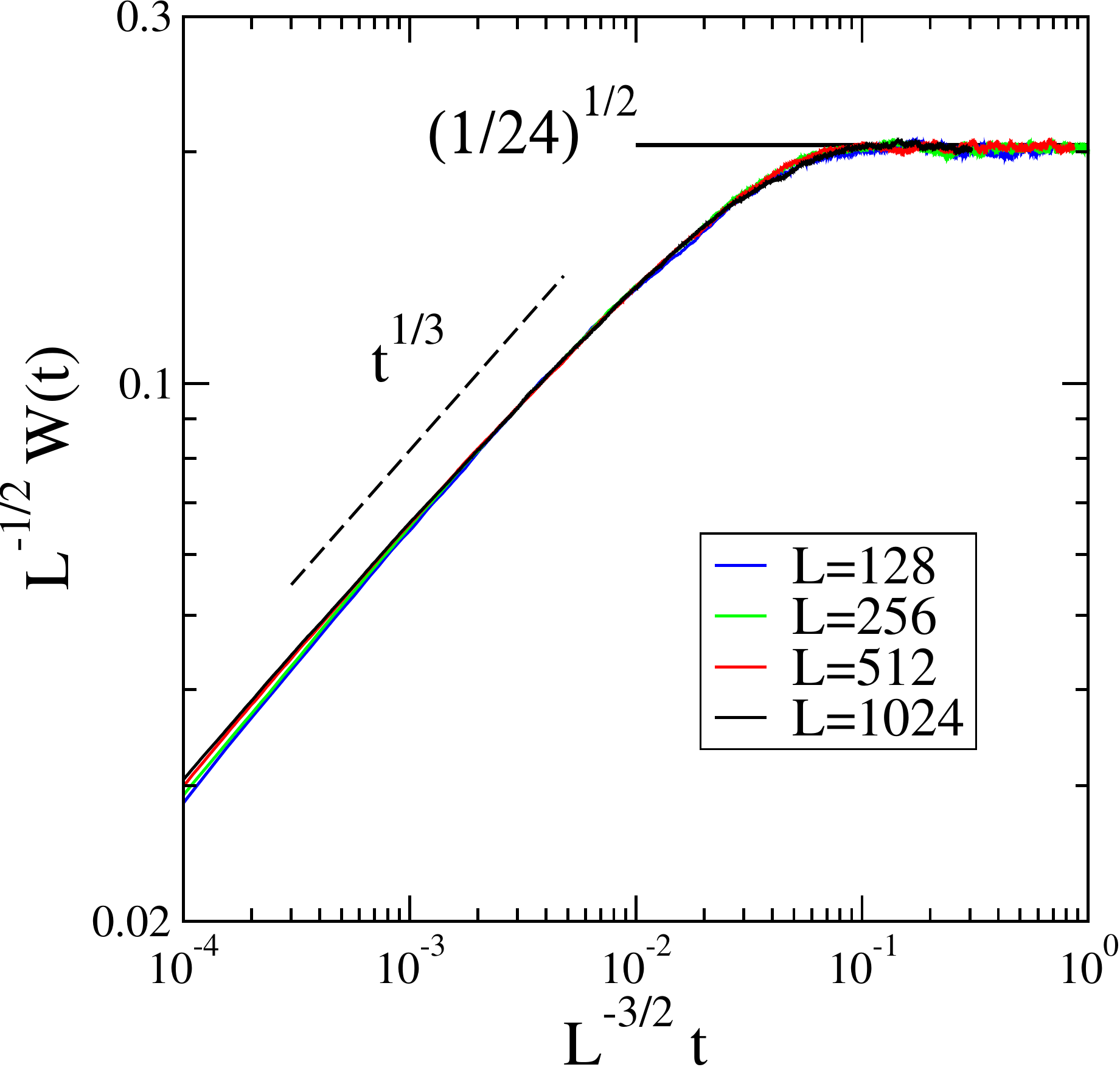}
\caption{Family--Vicsek scaling (\ref{fvscaling}) for the one-dimensional KPZ equation for different linear system sizes $L$.
A pseudospectral scheme with implicit-explicit time integration has been used to solve the KPZ equation (\ref{SKPZ}) with dimensionless coupling constant $g=8$. 
The data result from averaging 4800 runs with different realizations of the noise. 
The dashed line indicates a power-law increase with exponent $1/3$, whereas the full vertical line indicates the theoretical value $1/\sqrt{24}$ for the saturation width.
}
\label{fig1}
\end{figure}

Figure~\ref{fig1} shows data obtained with this pseudospectral scheme with implicit-explicit time integration. 
These data have been obtained with the integration time step $\delta t=0.0075$ and result from averaging $4800$ runs
with different realizations of the stochastic noise. 
In all cases we start with a flat surface at $t=0$. 
From the data we obtain the correct values for the different $(1+1)$-dimensional KPZ exponents, as verified in Fig.~\ref{fig1} through the Family--Vicsek scaling (\ref{fvscaling}) for systems of different sizes $L$. 
In addition, at saturation $L^{-1/2} W(t)$ is given by the theoretical value $1/\sqrt{24}$, indicated by the full horizontal line.

\section{Feedback-controlled KPZ equation}{\label{Sec 3}}

Due to its technological importance, the control of roughening processes has been the subject of previous studies \cite{Armaou00,Lou03,Lou06,Gomes15,Gomes17}. 
Focusing mostly on the Kuramoto--Sivashinsky equation, various schemes have been proposed that control the width during a growth process. 
In their interesting approach, Gomes \emph{et al.}~\cite{Gomes17} separate this equation into two parts, namely a deterministic non-linear and a stochastic linear equation. 
Control actuator functions are then used in a first step to stabilize the zero solution of the deterministic non-linear equation, whereas in a second step the roughness is controlled through the appropriate control of the stochastic linear equation.
In contrast, for the scheme presented in the following, we do not split the original growth equation into a linear and a non-linear equation.

Our starting point is the following controlled KPZ equation on a finite one-dimensional domain of length $L$ with periodic boundary conditions:
\begin{equation}
\frac{\partial h}{\partial{t}}=\frac{\partial^2{h}}{\partial{x^2}}+\frac{g}{2}\left|\frac{\partial{h}}{\partial{x}}\right|^2
+ \sum\limits_{n=-n_c, n \ne 0}^{n_c} b_n(x) u_n(t) +\eta(x,t)~,
\label{controlled_KPZ_eq}
\end{equation}
with the $2 n_c$ time-dependent inputs $u_n(t)$ that can be manipulated externally in order to change the dynamics in
a desired fashion, and the corresponding control actuator distribution functions $b_n(x)$, which fix how the action coming from the inputs $u_n(t)$ is distributed spatially. 
Commonly used control schemes include point actuator or periodic controls. 
In this work, we choose as actuators the periodic functions $b_n(x)=\exp \left(-i 2 \pi n x / L \right)$, and we treat the number of involved Fourier modes $n_c$ as a parameter that we vary and control.

As in the previous Section~II for the uncontrolled KPZ equation, we also use the pseudospectral method for the controlled KPZ equation, which yields as starting point the following coupled set of equations for the Fourier modes with wavenumber $k=2\pi m /L$:
\begin{eqnarray}
\partial_t {\hat{h}}(k,t) &&= -k^2 {\hat{h}}(k,t) + Q(k,{\hat{h}}) \nonumber\\
&&\quad +\sum_{n=-n_c, n \ne 0}^{n_c} \hat{b}_{n m}u_{n}(t) + \hat{\eta}(k,t)~,
\label{CF_KPZ}
\end{eqnarray}
with the same expression (\ref{NLKPZ}) for $Q(k,\hat{h})$, the Fourier transform of the KPZ non-linearity, whereas $\hat{b}_{n m}$ represents the term with $k = 2 \pi m /L$ from the Fourier series of the actuator function $b_n$.

Due to our choice for the actuator functions, $\hat{b}_{n m} = \delta_{nm}$, so that the $L+1$ coupled equations (\ref{CF_KPZ}) separate into two sets:
\begin{eqnarray}
\partial_t {\hat{h}}(k,t) &&= - \lambda_m {\hat{h}}(k,t) + \hat{\eta}(k,t) ~~ \mbox{if} ~ 0 < \left| m \right| \le n_c \label{modes_c} \\
\partial_t {\hat{h}}(k,t) &&= - k^2 {\hat{h}}(k,t) +Q(k,{\hat{h}}) + \hat{\eta}(k,t) \\
&&\qquad\qquad\qquad\qquad\quad \mbox{if} ~ L/2 \ge \left| m \right| >  n_c ~, \nonumber
\label{modes_u}
\end{eqnarray}
where for $0 < \left| m \right| \le n_c$ and $k=2 \pi m /L$ the eigenvalues of the controlled modes are given by
\begin{equation}
\lambda_k {\hat{h}}(k,t) = k^2 {\hat{h}}(k,t) - Q(k,{\hat{h}}) - u_k(t)~.
\end{equation}

In this setup the desired overall surface roughness is given by $W_d^2 = \lim_{t \rightarrow \infty} \langle W^2(t) \rangle$, with
\begin{eqnarray}
\langle W^2(t) \rangle & = & \sum\limits_{m=-L/2,m\ne 0}^{L/2} \langle {\hat{h}}^2(k,t) \rangle \nonumber \\
& = & \sum\limits_{m=-n_c,m\ne 0}^{n_c} \langle {\hat{h}}_c^2(k,t) \rangle + \sum\limits_{m=-L/2}^{n_c-1} \langle {\hat{h}}^2(k,t) \rangle \nonumber\\
&&+ \sum\limits_{m=n_c+1}^{L/2} \langle {\hat{h}}^2(k,t) \rangle ~,
\label{w2c}
\end{eqnarray}
where we use that $k = 2 \pi m /L$ with integer $m$.
For the controlled modes with $0 < |m| \le n_c$, we readily obtain from Eqs.~(\ref{modes_c}) the terms
\begin{equation} \label{h2_c}
\langle {\hat{h}}_c^2(k,t) \rangle = \frac{1}{2L \lambda_m} \left( 1 - e^{- 2 \lambda_m t} \right)~.
\end{equation}
We now assume that for the purpose of determining $\lambda_m$ for a given desired surface roughness, we can replace 
the solutions of (\ref{modes_u}) by those from the corresponding linearized equations. 
This yields for $|m| > n_c$ the terms
\begin{equation} \label{h2_u}
\langle {\hat{h}}^2(k,t) \rangle = \frac{1}{2 L k^2} \left( 1 - e^{- 2 k^2 t} \right)~.
\end{equation}
Inserting (\ref{h2_c}) and (\ref{h2_u}) into (\ref{w2c}) and taking the long-time limit $t \longrightarrow \infty$, we obtain
\begin{equation}
W_d^2 = \sum\limits_{m=-n_c,m\ne 0}^{n_c} \frac{1}{2L \lambda_m} + \frac{L}{4 \pi^2} \sum\limits_{m=n_c+1}^{L/2} \frac{1}{m^2} ~,
\end{equation}
and, after choosing the same value $\lambda_m = \lambda$ for each $m$:
\begin{equation}
\lambda = \frac{n_c}{L} \frac{1}{W_d^2 - \frac{L}{4 \pi^2} \sum\limits_{m=n_c+1}^{L/2} \frac{1}{m^2}} ~.
\end{equation}
It is this value of $\lambda$ that determines our manipulated inputs
\begin{equation} \label{uk}
u_k(t) = (k^2 - \lambda) {\hat{h}}(k,t) - Q(k, {\hat{h}})
\end{equation}
in the controlled KPZ equations (\ref{CF_KPZ}).

The integration scheme outlined in Section~\ref{Sec 2} can be applied in a straightforward manner to integrate the controlled KPZ equation (\ref{CF_KPZ}) with the time-dependent inputs (\ref{uk}).

\section{Numerical Results} {\label{Sec 4}}

\begin{figure}
 \centering \includegraphics[width=0.6\columnwidth,clip=true]{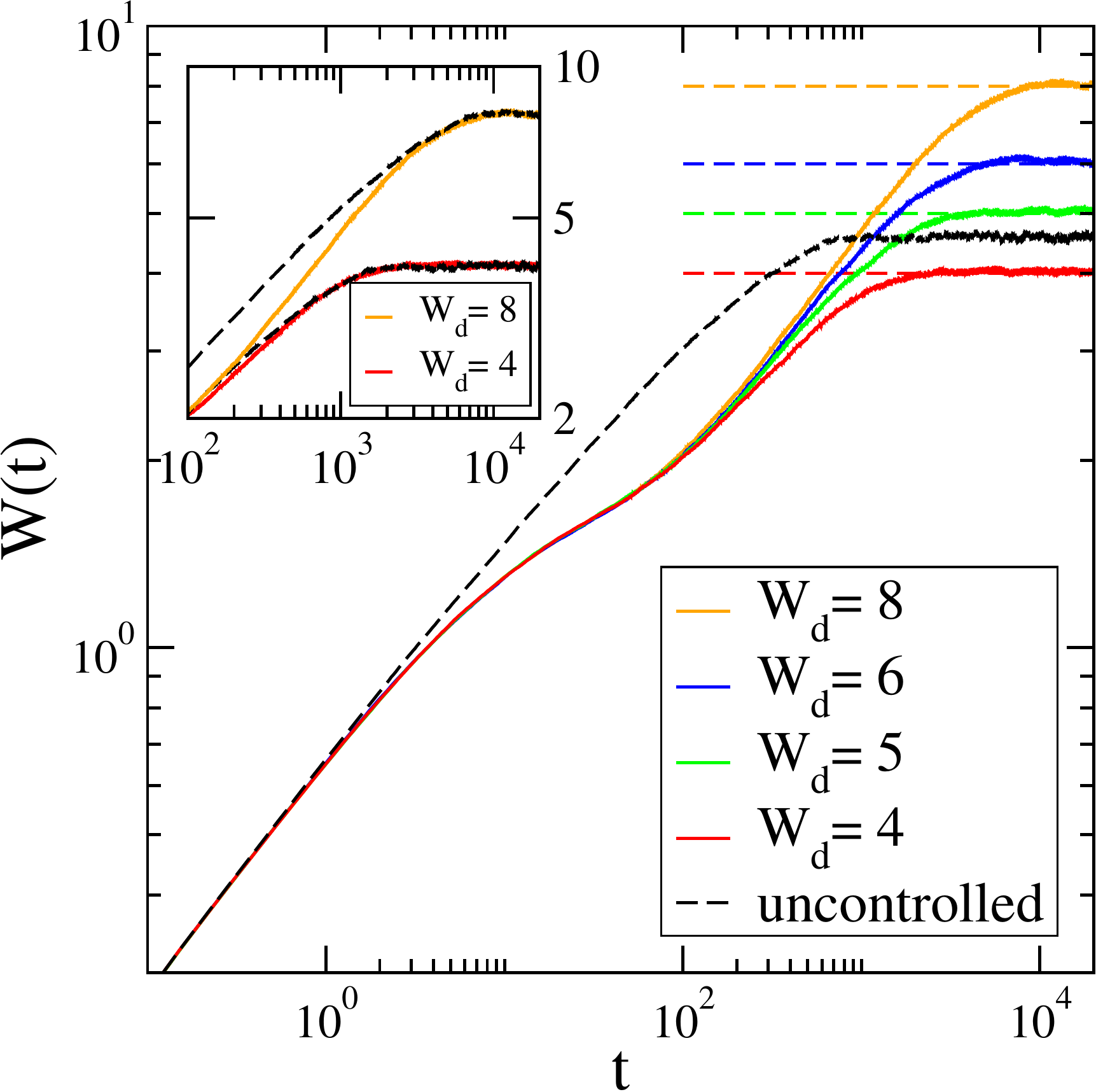}
\caption{Time-dependent surface width as obtained from the controlled KPZ equation (\ref{controlled_KPZ_eq}) with $n_c=6$ for different desired saturation widths $W_d$. 
The system parameters here are $L=512$, $g=8$, and $\delta t = 0.0075$. 
Also displayed for comparison are data for the uncontrolled case with the same parameter values. 
The inset shows that the approach to the steady state is similar for both the uncontrolled and controlled growth processes. The dashed black lines represent the same data as in the main figure for the uncontrolled system, only shifted so that the steady-state plateaus coincide with those for the controlled data.
The data for the controlled cases are an average over at least $8000$ runs with different realizations of the noise.
}
\label{fig2}
\end{figure}

Previously proposed schemes to control the surface width in a growth process \cite{Armaou00,Lou03,Lou06,Gomes15,Gomes17} effectively aim at suppressing the non-linearities in the growth equations. As a result, the time-dependent controlled surface width loses completely the scaling properties that characterize a non-linear growth universality class such as KPZ. 
For example, the scheme proposed in Ref.~\cite{Gomes17} results in a time-dependent increase of the width with an exponent that is independent of the type of non-linearity in the uncontrolled equation. 
As we show in the following, this is different in our scheme described in Section~\ref{Sec 3}, which results in non-trivial scaling behavior of the controlled surface width.

We start the discussion of our numerical results with Fig. \ref{fig2} that shows the time evolution of the width $W(t)$ for different target saturation values $W_d$ and a fixed number of control actuators, $n_c=6$. Inspection of the main
panel reveals the appearance of several distinct temporal regimes during this controlled stochastic non-linear dynamics.
In the very early time regime, the controlled surface width follows the width of the uncontrolled KPZ surface, indicated by the dashed black line in the figure, before the growth of the width is slowed down dramatically. 
This remarkable behavior results from the interplay between non-linear dynamics for the majority of the Fourier modes with linear dynamics for the remaining $2 n_c$ modes. 
This is followed by a regime of faster surface growth where the curves approaching different $W_d$ finally separate prior to saturating at their target values. 
As depicted in the inset, this final approach to saturation looks very similar to the crossover to the steady state for the uncontrolled KPZ equation.

We emphasize that the behavior shown in Fig. \ref{fig2} differs drastically from that obtained from the scheme in Ref.~\cite{Gomes17}, where the dynamics is separated into a linear and a non-linear equation, with the solution of the non-linear equation set to zero at each integration step. 
Consequently, the linear evolution dominates the roughness, yielding scaling exponent values that are independent of the specific nature of the non-linearity. 
In contrast, in our scheme the non-linearity continues to impact the evolution of the roughness significantly until ultimately the targeted surface width has been reached.

\begin{figure}
 \centering \includegraphics[width=0.6\columnwidth,clip=true]{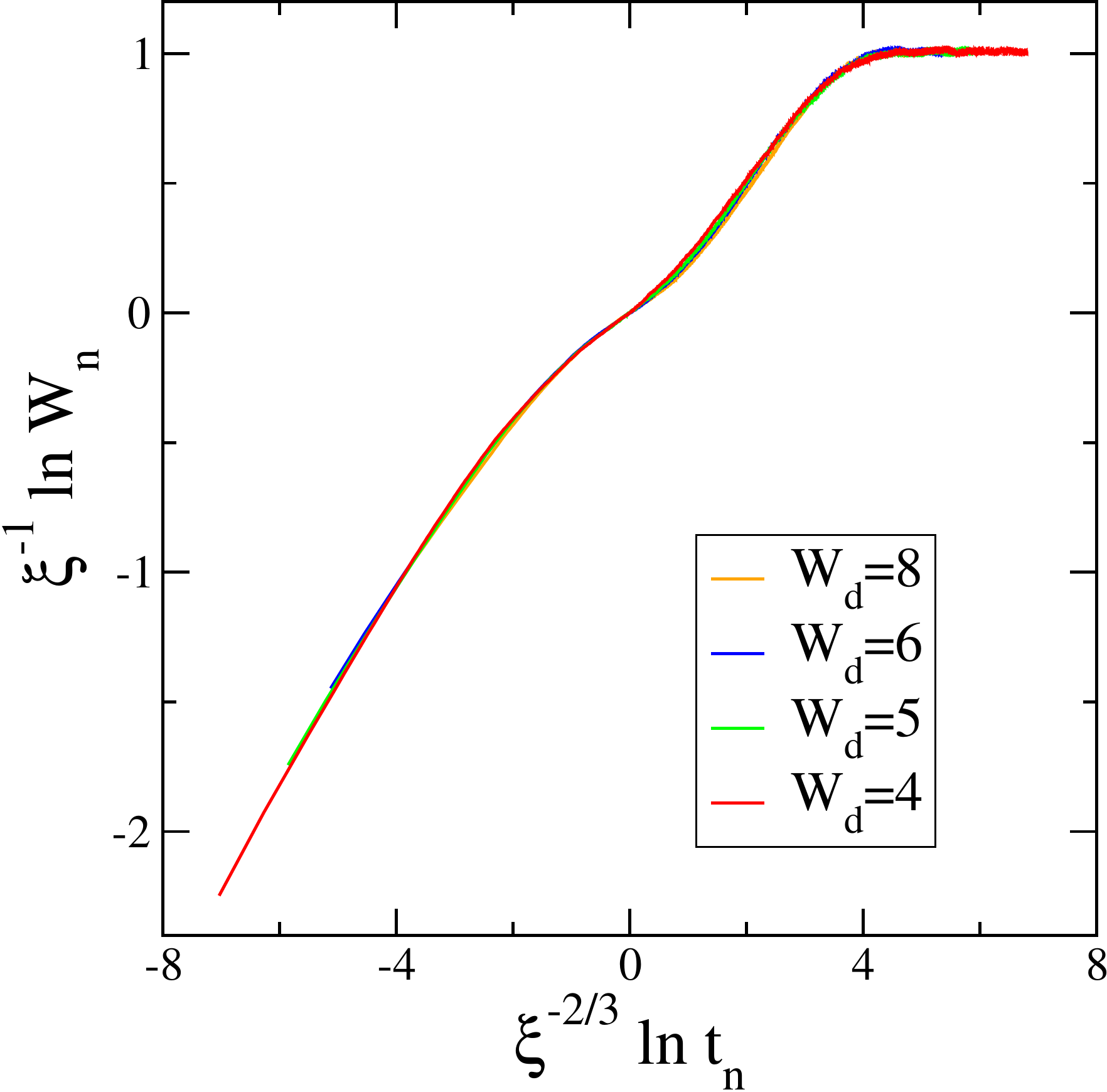}
\caption{Generalized scaling of the time-dependent surface width obtained from the controlled KPZ equation for different saturation widths $W_d$, see Fig.~\ref{fig2} for the unscaled data. 
The system parameters are $L=512$, $g=8$, $n_c=6$, and $\delta t = 0.0075$. 
The data are the average of an ensemble of at least $8000$ runs with different realizations of the noise.
}
\label{fig3}
\end{figure}

The variety of multiple temporal regimes in the controlled growth process is reminiscent of the appearance of different regimes in competitive growth models \cite{Chou09}. 
Examples include the (random deposition / random deposition with surface relaxation (RD / RDSR) model \cite{Horowitz01}, where the deposition of particles happens with probability $p$ respectively $1-p$ following the RDSR respectively RD rules, which results in a crossover between these two distinct regimes before the system settles into its steady state; or the restricted solid-on-solid (RSOS) model \cite{Kim89} that displays a crossover from a random-deposition to a KPZ regime, followed by the final crossover to saturation. 
For this type of competitive growth systems, a generalized scaling law has been demonstrated \cite{Chou09}. 
Using rescaled variables $W_n = W/W_1$ and $t_n = t/t_1$ where $t_1$ and $W_1$ are the (system parameter-dependent) time and surface height at the crossover location between the first two regimes (\emph{e.g.}, between random deposition and KPZ for the RSOS model), the following scaling relation yields a complete data collapse for such competitive growth models:
\begin{equation} \label{chou1}
\ln(W_n) / \xi  = F \left( \ln(t_n)/ \xi \right) ,
\end{equation}
where $\xi = \ln (W_2/W_1)$ is the logarithm of the ratio between the steady-state value $W_2$ and the surface width $W_1$ at the crossover separating the two early-time regimes, while $F$ represents a scaling function that only depends on the scaled variable $\ln(t_n)/ \xi$.

The behavior of the controlled surface width in our scheme is in fact more complex than those encountered in competitive growth models, displaying four distinctive dynamical regimes. 
For that reason we consider the following generalization of Eq.~(\ref{chou1}):
\begin{equation} \label{chou2}
\ln(W_n) / \xi^a  = F \left( \ln(t_n)/ \xi^b \right) ,
\end{equation}
where the exponents $a$ and $b$ should be determined from optimal data collapse. 
Whereas in our case $W_2$ is given by the target saturation width $W_d$, for $t_1$ and $W_1$ we choose the values of $t$ and $W$ at the point where the data for different $W_d$ start to separate. 
We have checked that the values of the exponents $a$ and $b$ we obtain from the best data collapse do not depend on this specific choice as long as $t_1$ and $W_1$ are taken inside the region where the change from slower to faster growth takes place.
We then obtain the excellent data collapse shown in Fig.~\ref{fig3} with the values $a=1$ and $b=2/3$ for the exponents. 
At this point we cannot provide an explanation why the exponent $b$, which equals $1$ in the competitive growth models \cite{Chou09}, takes on the value $2/3$ during controlled KPZ surface growth. 
It is worth noting, though, that we obtain the same value for this exponent when we change the number of controlled Fourier modes.

\begin{figure}
 \centering \includegraphics[width=0.6\columnwidth,clip=true]{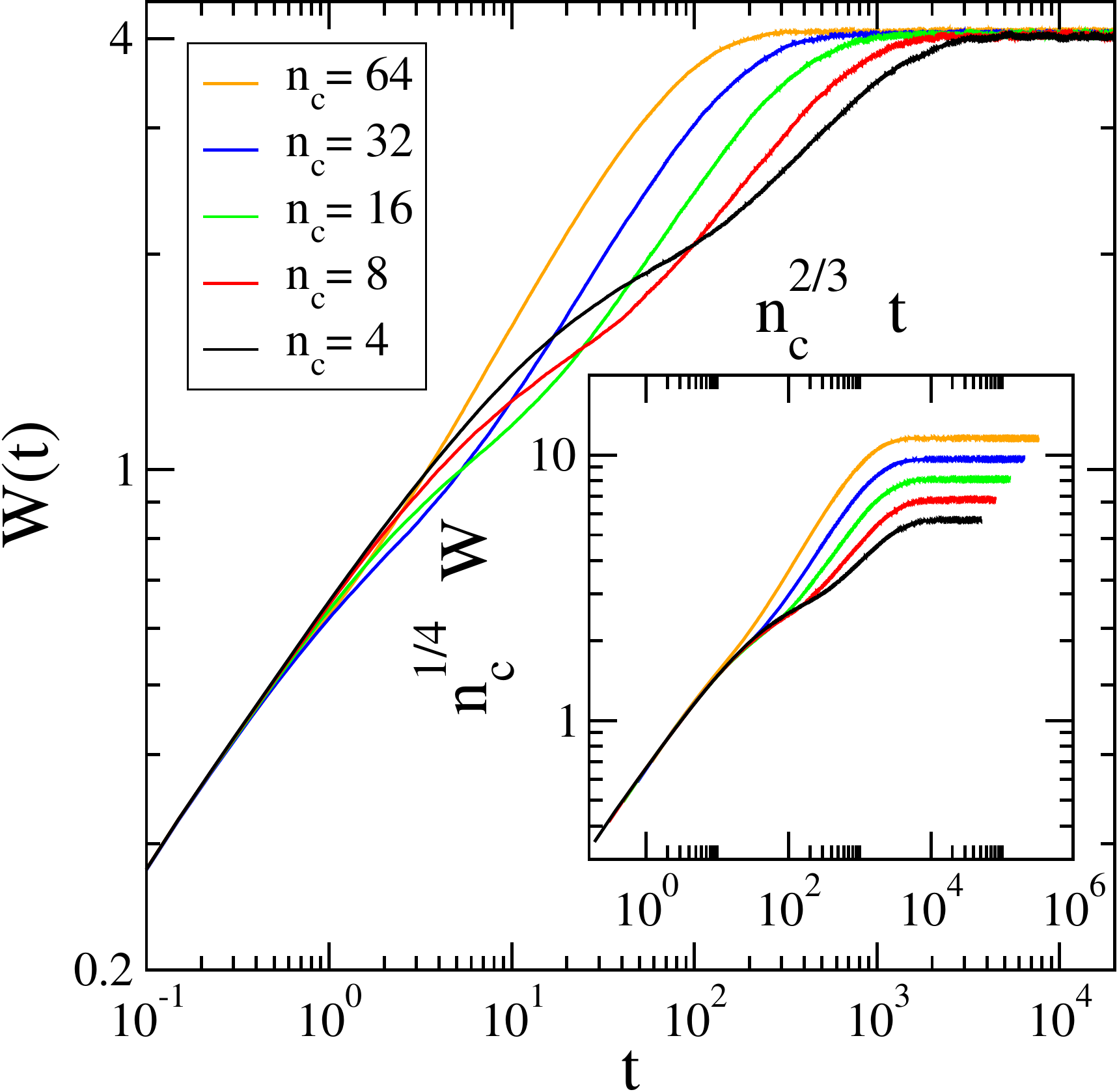}
\caption{Time-dependent surface width for various numbers $2 n_c$ of controlled modes when the target saturation width is chosen to be $W_d=4$.
The scaled data shown in the inset reveal a systematic crossover behavior. 
The system parameters are $L=512$, $g=8$, and $\delta t = 0.0075$. 
The data are the average of an ensemble of at least $8000$ runs with different realizations of the noise.
}
\label{fig4}
\end{figure}

Figure~\ref{fig4} illustrates how the surface width $W(t)$ for a fixed target saturation width $W_d$ depends on the number of controlled modes $2 n_c$. 
The most notable effect is that the crossover from the slow- to the fast-growth regime happens earlier for larger $n_c$.
In addition, the exponent governing the quick growth before the crossover to saturation increases in magnitude, \emph{e.g.}, taking the value $0.36$ for $n_c = 16$, whereas for $n_c=64$ its value is $0.41$. 
This increase towards the value $0.5$ reflects the fact that for a larger number of controlled Fourier modes, the contribution of the linear dynamics becomes enhanced. 
As revealed in the inset, appropriate rescaling of time and surface width allows the collapse of the data obtained for various $n_c$ on a single master curve in the intermediate temporal regime prior to the inception of the fast growth and saturation. 
Again, this remarkable data collapse is achieved with non-trivial scaling exponents, whose actual values we cannot yet predict or explain.
 
\begin{figure}
 \centering \includegraphics[width=0.6\columnwidth,clip=true]{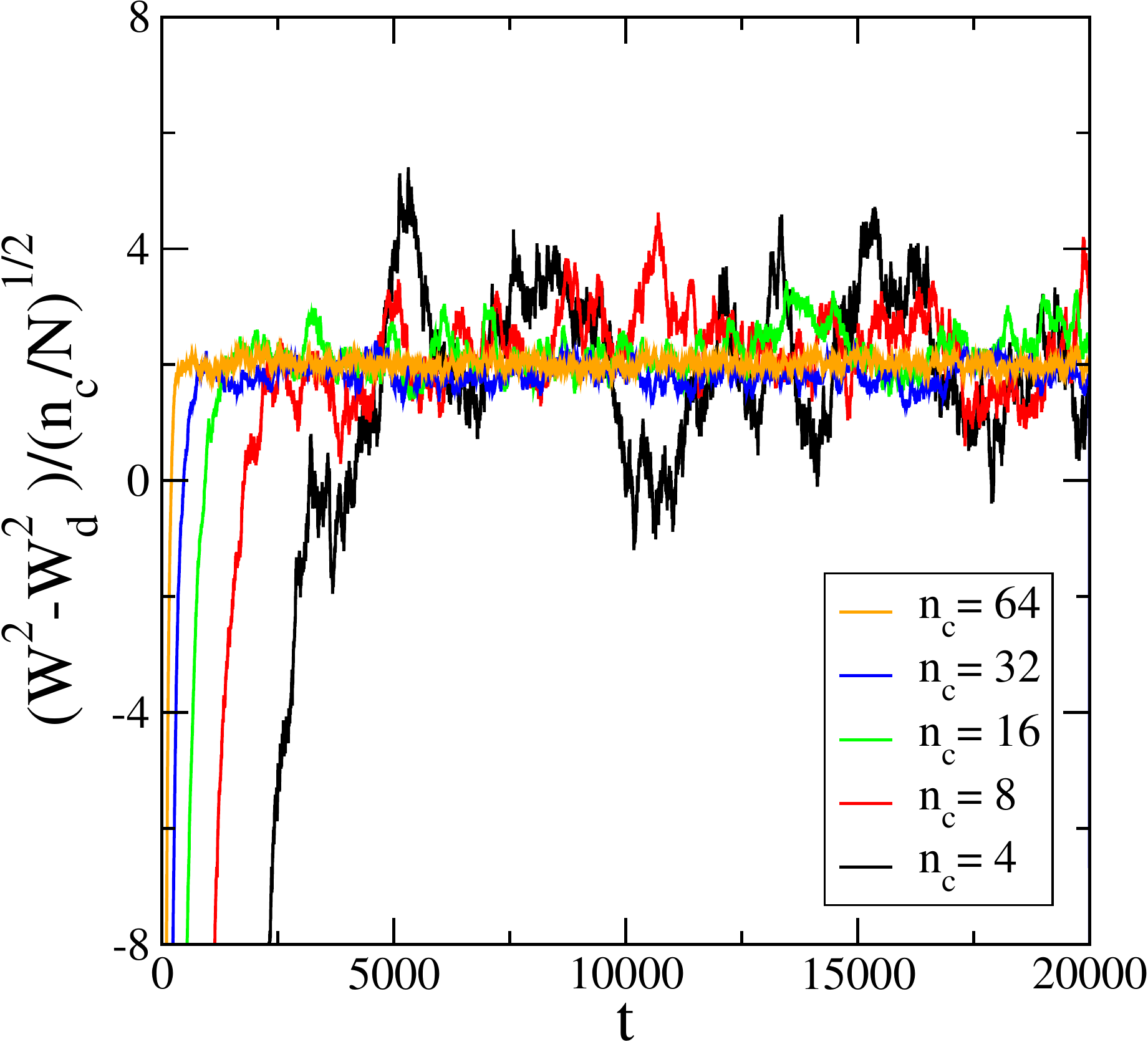}
\caption{Scaling of the deviations from the target saturation with $W_d=4$, where data for systems with different numbers $2 n_c$ of controlled modes are shown. 
The system parameters are $L=512$, $g=8$, and $\delta t = 0.0075$. 
The data result from an ensemble average over at least $8000$ runs with different realizations of the noise.
}
\label{fig5}
\end{figure}

Our scheme allows us to control the final roughness of a stochastically growing surface while retaining some of the complexity of the original non-linear growth process, but of course it is still an approximate method. 
The most consequential step is the omission of non-linear contributions that results in the expression (\ref{h2_u}) for the uncontrolled modes. 
As a result of this approximation, the steady-state surface width in fact exhibits small but systematic deviations from the target saturation width $W_d$. 
These deviations can be regarded as forming three different regimes depending on the number of control actuators. 
In the first regime, with $n_c/N < 1/4$, this deviation does not depend on the strength $g$ of the non-linearity. 
Instead, a dependence on $n_c/N$ in the form of a simple scaling behavior is observed after some transient time:
\begin{equation} \label{width_dev}
W^2 - W^2_d = \left( \frac{n_c}{N} \right)^{1/2} Z(W_d)~,
\end{equation}
as shown in Fig.~\ref{fig5}. For larger values of $n_c/N$, the strength of the non-linearity does affect the amount the steady-state surface width deviates from the target value. 
Moreover, while a scaling ansatz like (\ref{width_dev}) still holds to a good approximation, the scaling exponent is found to deviate from the value $1/2$, increasing for $1/4 < n_c/N < 1/2$, while decreasing in the range $1/2 < n_c/N < 1$. 
When $n_c/N \longrightarrow 1$, the dynamics becomes entirely linear and no deviations from the target value are observed anymore.

Additional insights can be gained from the skewness of the height probability distribution,
\begin{equation} \label{eq_gamma}
\gamma = \left< \left( \frac{h - \langle N \rangle}{W^2} \right)^3 \right>~.
\end{equation}
Whereas for the uncontrolled KPZ equation the surface fluctuations are Gaussian with vanishing skewness, for the controlled KPZ growth process deviations from Gaussianity are to be expected. 
These deviations will of course disappear for $n_c/N \longrightarrow 1$, when the dynamics becomes fully linear.

\begin{figure}
 \centering \includegraphics[width=0.6\columnwidth,clip=true]{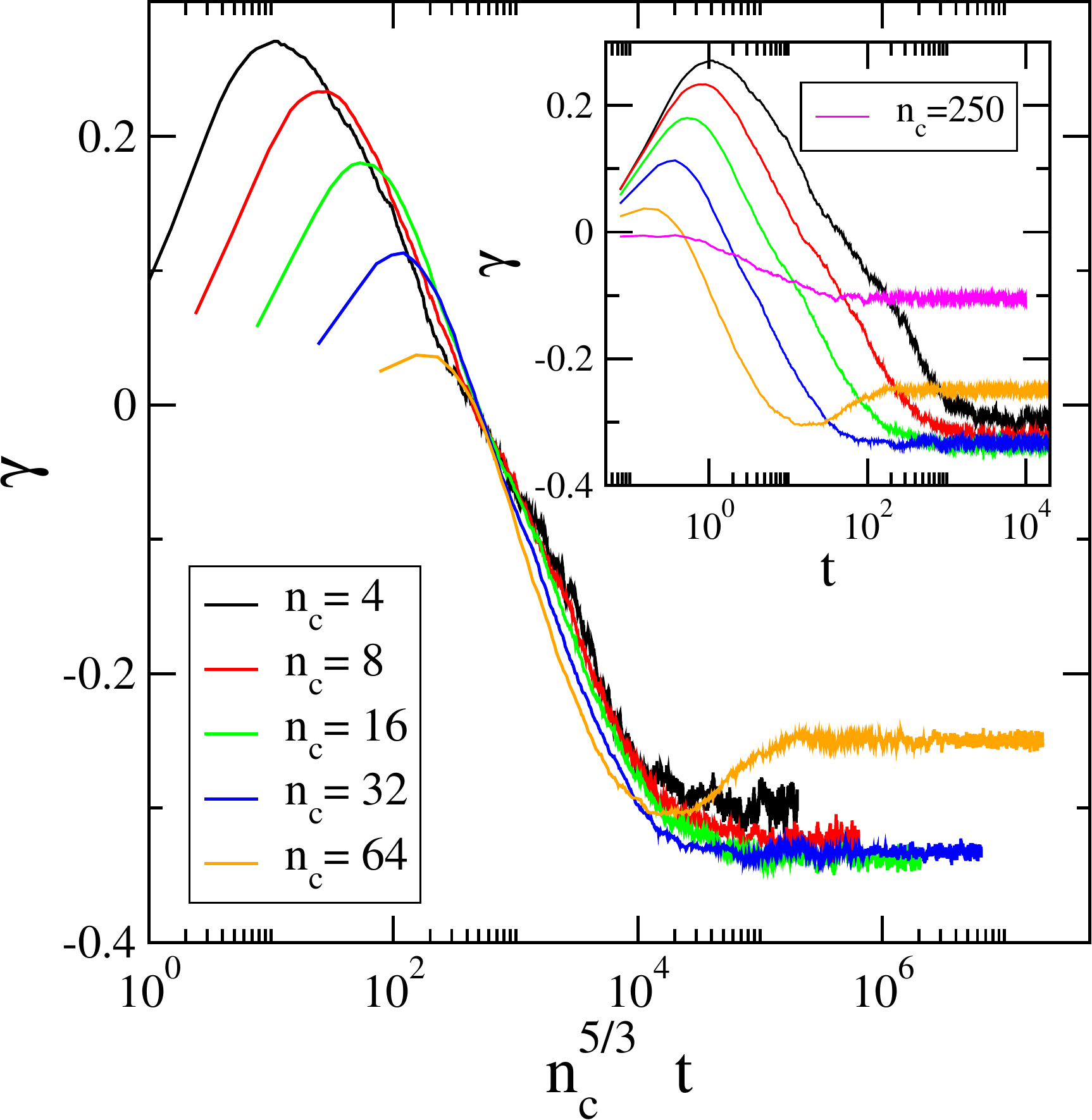}
\caption{Data collapse when plotting the skewness $\gamma$ as a function of rescaled times $n_c^{5/3} t$. 
Data for systems with different numbers $2 n_c$ of controlled modes are shown; the inset depicts the unscaled data. 
The system parameters are $L=512$, $g=8$, $W_d=4$, and $\delta t = 0.0075$. 
The data result from an ensemble average over at least $8000$ runs with different realizations of the noise.
}
\label{fig6}
\end{figure}

The time-dependent skewness $\gamma(t)$ displays interesting features, as illustrated in Fig.~\ref{fig6}. 
We first note, see the inset, that for a small number of controlled modes with $n_c/N < 1/4$, the skewness changes sign as time progresses, being positive at early times before subsequently becoming negative and approaching a plateau value that characterizes the steady state. 
As shown in the main panel, an approximate data collapse is obtained in the regime $n_c/N < 1/4$ if time is rescaled with $n_c$ according to $n_c^{5/3} t$. 
For larger values of $n_c/N$ the pattern changes, as demonstrated in the inset of Fig.~\ref{fig6}: 
The skewness now is negative from the start, no acceptable scaling can be achieved anymore, and the magnitude of its steady-state value increases and approaches zero for the limiting case $n_c/N \longrightarrow 1$.
Comparison of the behaviors of $\gamma(t)$ in Fig.~\ref{fig6} and of the deviations from the target surface width in 
Fig.~\ref{fig5} allows us to establish a direct connection of the properties of the height probability distribution with both the time evolution and the steady-state properties of the surface width obtained from the non-linear feedback controlled KPZ equation.

\section{Conclusion}{\label{Sec 5}}

Targeting specific macroscopic properties through external feedback control for complex cooperative non-linear stochastic dynamics constitutes a largely open problem which entails fundamental issues that are both important from a basic theoretical as well as application viewpoint.
For example, controlling the overall roughness of a surface during a non-linear stochastic growth process is of obvious technological importance. 
In this paper we have presented an approach that crucially differs from previously discussed control schemes. 
Whereas other schemes basically aim at suppressing as much as possible the non-linear character of the underlying growth process in order to force a desired width on the fluctuating surface, our protocol, which accounts for the coupling between
controlled and uncontrolled Fourier modes, allows us to target the desired surface roughness while maintaining many non-trivial scaling aspects of the underlying non-linear growth process.

In this work we focus on the one-dimensional KPZ equation as a paradigmatic example, but our scheme can of course be
generalized to other non-linear stochastic growth processes like the Kuramoto--Sivashinsky equation. 
Using a robust pseudospectral third-order Runge-Kutta scheme with mixed time integration, we have explored both the transient and steady-state surface fluctuations upon choosing periodic functions as actuators. 
A particular emphasis of our investigation has been the impact of changing the number of controlled Fourier modes.

Targeting with our scheme to a desired steady-state value the overall interface width yields intriguingly complex temporal behavior. 
We may even force the surface to have either a smaller or larger roughness than that of the uncontrolled growth process with the same system parameters. 
Interestingly, this can be achieved by retaining much of the scale-invariant complex dynamics of the underlying KPZ growth process, as revealed by the non-trivial scaling properties of our data.
An important parameter is the number of actuator functions used in the control scheme, as it determines the mix and relative weight between non-linear and linear dynamics. 
Various different crossover regimes can be identified, as increasing the number of actuator functions pushes the system's dynamics further away from the uncontrolled KPZ dynamics and closer to a fully linear relaxation kinetics.
These different regimes are also encoded in the height probability distribution, as we have shown by investigating its skewness.

Our work provides an obvious starting point for numerous possible future investigations. 
We have focused here entirely on a numerical study, but it will be worth exploring whether some of the observed scaling properties, specifically, the values of the various crossover scaling exponents, can be understood through an analytical treatment. 
The scheme presented here can, as already mentioned, be applied and extended to other situations, including the experimentally more relevant $(2+1)$-dimensional KPZ equation. 
Finally, other related control schemes can be imagined and constructed that might display different properties in the transient and / or steady-state regimes. 


\acknowledgments
Priyanka would like to thank James Stidham and Shannon Serrao for fruitful discussions, and Vicky Verma for introducing her to the numerical integration scheme.
Research was sponsored by the Army Research Office and was accomplished under Grant Number \textbf{W911NF-17-1-0156}. The views and 
conclusions contained in this document are those of the authors and should not be interpreted as representing the official policies, 
either expressed or implied, of the Army Research Office or the U.S. Government. The U.S. Government is authorized to reproduce and 
distribute reprints for Government purposes notwithstanding any copyright notation herein.

%


\begin{thebibliography}{25}%
\makeatletter
\providecommand \@ifxundefined [1]{%
 \@ifx{#1\undefined}
}%
\providecommand \@ifnum [1]{%
 \ifnum #1\expandafter \@firstoftwo
 \else \expandafter \@secondoftwo
 \fi
}%
\providecommand \@ifx [1]{%
 \ifx #1\expandafter \@firstoftwo
 \else \expandafter \@secondoftwo
 \fi
}%
\providecommand \natexlab [1]{#1}%
\providecommand \enquote  [1]{``#1''}%
\providecommand \bibnamefont  [1]{#1}%
\providecommand \bibfnamefont [1]{#1}%
\providecommand \citenamefont [1]{#1}%
\providecommand \href@noop [0]{\@secondoftwo}%
\providecommand \href [0]{\begingroup \@sanitize@url \@href}%
\providecommand \@href[1]{\@@startlink{#1}\@@href}%
\providecommand \@@href[1]{\endgroup#1\@@endlink}%
\providecommand \@sanitize@url [0]{\catcode `\\12\catcode `\$12\catcode
  `\&12\catcode `\#12\catcode `\^12\catcode `\_12\catcode `\%12\relax}%
\providecommand \@@startlink[1]{}%
\providecommand \@@endlink[0]{}%
\providecommand \url  [0]{\begingroup\@sanitize@url \@url }%
\providecommand \@url [1]{\endgroup\@href {#1}{\urlprefix }}%
\providecommand \urlprefix  [0]{URL }%
\providecommand \Eprint [0]{\href }%
\providecommand \doibase [0]{https://doi.org/}%
\providecommand \selectlanguage [0]{\@gobble}%
\providecommand \bibinfo  [0]{\@secondoftwo}%
\providecommand \bibfield  [0]{\@secondoftwo}%
\providecommand \translation [1]{[#1]}%
\providecommand \BibitemOpen [0]{}%
\providecommand \bibitemStop [0]{}%
\providecommand \bibitemNoStop [0]{.\EOS\space}%
\providecommand \EOS [0]{\spacefactor3000\relax}%
\providecommand \BibitemShut  [1]{\csname bibitem#1\endcsname}%
\let\auto@bib@innerbib\@empty
\bibitem [{\citenamefont {Kardar}\ \emph {et~al.}(1986)\citenamefont {Kardar},
  \citenamefont {Parisi},\ and\ \citenamefont {Zhang}}]{Kardar86}%
  \BibitemOpen
  \bibfield  {author} {\bibinfo {author} {\bibfnamefont {M.}~\bibnamefont
  {Kardar}}, \bibinfo {author} {\bibfnamefont {G.}~\bibnamefont {Parisi}},\
  and\ \bibinfo {author} {\bibfnamefont {Y.~C.}\ \bibnamefont {Zhang}},\
  }\bibfield  {title} {\bibinfo {title} {Dynamic scaling of growing
  interfaces},\ }\href@noop {} {\bibfield  {journal} {\bibinfo  {journal}
  {Phys. Rev. Lett.}\ }\textbf {\bibinfo {volume} {56}},\ \bibinfo {pages}
  {889} (\bibinfo {year} {1986})}\BibitemShut {NoStop}%
\bibitem [{\citenamefont {Kuramoto}(1978)}]{Kuramoto78}%
  \BibitemOpen
  \bibfield  {author} {\bibinfo {author} {\bibfnamefont {Y.}~\bibnamefont
  {Kuramoto}},\ }\bibfield  {title} {\bibinfo {title} {Diffusion-induced chaos
  in reaction systems},\ }\href@noop {} {\bibfield  {journal} {\bibinfo
  {journal} {Prog. Theor. Phys. Supp.}\ }\textbf {\bibinfo {volume} {64}},\
  \bibinfo {pages} {346} (\bibinfo {year} {1978})}\BibitemShut {NoStop}%
\bibitem [{\citenamefont {Sivashinsky}(1980)}]{Sivashinsky80}%
  \BibitemOpen
  \bibfield  {author} {\bibinfo {author} {\bibfnamefont {G.~I.}\ \bibnamefont
  {Sivashinsky}},\ }\bibfield  {title} {\bibinfo {title} {On flame propagation
  under conditions of stoichiometry},\ }\href {https://doi.org/10.1137/0139007}
  {\bibfield  {journal} {\bibinfo  {journal} {SIAM J. Appl. Math.}\ }\textbf
  {\bibinfo {volume} {39}},\ \bibinfo {pages} {67} (\bibinfo {year}
  {1980})}\BibitemShut {NoStop}%
\bibitem [{\citenamefont {Armaou}\ and\ \citenamefont
  {Christofides}(2000)}]{Armaou00}%
  \BibitemOpen
  \bibfield  {author} {\bibinfo {author} {\bibfnamefont {A.}~\bibnamefont
  {Armaou}}\ and\ \bibinfo {author} {\bibfnamefont {P.~D.}\ \bibnamefont
  {Christofides}},\ }\bibfield  {title} {\bibinfo {title} {Feedback control of
  the {K}uramoto--{S}ivashinsky equation},\ }\href@noop {} {\bibfield  {journal}
  {\bibinfo  {journal} {Physica D: Nonlinear Phenomena}\ }\textbf {\bibinfo
  {volume} {137}},\ \bibinfo {pages} {49 } (\bibinfo {year}
  {2000})}\BibitemShut {NoStop}%
\bibitem [{\citenamefont {Lou}\ and\ \citenamefont
  {Christofides}(2003)}]{Lou03}%
  \BibitemOpen
  \bibfield  {author} {\bibinfo {author} {\bibfnamefont {Y.}~\bibnamefont
  {Lou}}\ and\ \bibinfo {author} {\bibfnamefont {P.~D.}\ \bibnamefont
  {Christofides}},\ }\bibfield  {title} {\bibinfo {title} {Optimal
  actuator/sensor placement for nonlinear control of the
  {K}uramoto-{S}ivashinsky equation},\ }\href@noop {} {\bibfield  {journal}
  {\bibinfo  {journal} {IEEE Transactions on Control Systems Technology}\
  }\textbf {\bibinfo {volume} {11}},\ \bibinfo {pages} {737} (\bibinfo {year}
  {2003})}\BibitemShut {NoStop}%
\bibitem [{\citenamefont {Lou}\ and\ \citenamefont
  {Christofides}(2006)}]{Lou06}%
  \BibitemOpen
  \bibfield  {author} {\bibinfo {author} {\bibfnamefont {Y.}~\bibnamefont
  {Lou}}\ and\ \bibinfo {author} {\bibfnamefont {P.~D.}\ \bibnamefont
  {Christofides}},\ }\bibfield  {title} {\bibinfo {title} {Nonlinear feedback
  control of surface roughness using a stochastic PDE:  design and
  application to a sputtering process},\ }\href@noop {} {\bibfield  {journal}
  {\bibinfo  {journal} {Ind. \& Eng. Chem. Res.}\ }\textbf {\bibinfo {volume}
  {45}},\ \bibinfo {pages} {7177} (\bibinfo {year} {2006})}\BibitemShut
  {NoStop}%
\bibitem [{\citenamefont {Gomes}\ \emph {et~al.}(2015)\citenamefont {Gomes},
  \citenamefont {Pradas}, \citenamefont {Kalliadasis}, \citenamefont
  {Papageorgiou},\ and\ \citenamefont {Pavliotis}}]{Gomes15}%
  \BibitemOpen
  \bibfield  {author} {\bibinfo {author} {\bibfnamefont {S.~N.}\ \bibnamefont
  {Gomes}}, \bibinfo {author} {\bibfnamefont {M.}~\bibnamefont {Pradas}},
  \bibinfo {author} {\bibfnamefont {S.}~\bibnamefont {Kalliadasis}}, \bibinfo
  {author} {\bibfnamefont {D.~T.}\ \bibnamefont {Papageorgiou}},\ and\ \bibinfo
  {author} {\bibfnamefont {G.~A.}\ \bibnamefont {Pavliotis}},\ }\bibfield
  {title} {\bibinfo {title} {Controlling spatiotemporal chaos in active
  dissipative-dispersive nonlinear systems},\ }\href@noop {} {\bibfield
  {journal} {\bibinfo  {journal} {Phys. Rev. E}\ }\textbf {\bibinfo {volume}
  {92}},\ \bibinfo {pages} {022912} (\bibinfo {year} {2015})}\BibitemShut
  {NoStop}%
\bibitem [{\citenamefont {Gomes}\ \emph {et~al.}(2017)\citenamefont {Gomes},
  \citenamefont {Kalliadasis}, \citenamefont {Papageorgiou}, \citenamefont
  {Pavliotis},\ and\ \citenamefont {Pradas}}]{Gomes17}%
  \BibitemOpen
  \bibfield  {author} {\bibinfo {author} {\bibfnamefont {S.~N.}\ \bibnamefont
  {Gomes}}, \bibinfo {author} {\bibfnamefont {S.}~\bibnamefont {Kalliadasis}},
  \bibinfo {author} {\bibfnamefont {D.~T.}\ \bibnamefont {Papageorgiou}},
  \bibinfo {author} {\bibfnamefont {G.}~\bibnamefont {Pavliotis}},\ and\
  \bibinfo {author} {\bibfnamefont {M.}~\bibnamefont {Pradas}},\ }\bibfield
  {title} {\bibinfo {title} {Controlling roughening processes in the stochastic
  {Kuramoto}--{Sivashinsky} equation},\ }\href@noop {} {\bibfield  {journal}
  {\bibinfo  {journal} {Physica D: Nonlinear Phenomena}\ }\textbf {\bibinfo
  {volume} {348}},\ \bibinfo {pages} {33 } (\bibinfo {year}
  {2017})}\BibitemShut {NoStop}%
\bibitem [{\citenamefont {Family}\ and\ \citenamefont
  {Vicsek}(1985)}]{Family85}%
  \BibitemOpen
  \bibfield  {author} {\bibinfo {author} {\bibfnamefont {F.}~\bibnamefont
  {Family}}\ and\ \bibinfo {author} {\bibfnamefont {T.}~\bibnamefont
  {Vicsek}},\ }\bibfield  {title} {\bibinfo {title} {Scaling of the active zone
  in the {Eden} process on percolation networks and the ballistic deposition
  model},\ }\href@noop {} {\bibfield  {journal} {\bibinfo  {journal} {J. Phys
  A: Math. and Gen.}\ }\textbf {\bibinfo {volume} {18}},\ \bibinfo {pages}
  {L75} (\bibinfo {year} {1985})}\BibitemShut {NoStop}%
\bibitem [{\citenamefont {Chou}\ and\ \citenamefont
  {Pleimling}(2009)}]{Chou09}%
  \BibitemOpen
  \bibfield  {author} {\bibinfo {author} {\bibfnamefont {Y.-L.}\ \bibnamefont
  {Chou}}\ and\ \bibinfo {author} {\bibfnamefont {M.}~\bibnamefont
  {Pleimling}},\ }\bibfield  {title} {\bibinfo {title} {Parameter-free scaling
  relation for nonequilibrium growth processes},\ }\href@noop {} {\bibfield
  {journal} {\bibinfo  {journal} {Phys. Rev. E}\ }\textbf {\bibinfo {volume}
  {79}},\ \bibinfo {pages} {051605} (\bibinfo {year} {2009})}\BibitemShut
  {NoStop}%
\bibitem [{\citenamefont {Forster}\ \emph {et~al.}(1977)\citenamefont
  {Forster}, \citenamefont {Nelson},\ and\ \citenamefont
  {Stephen}}]{Forster77}%
  \BibitemOpen
  \bibfield  {author} {\bibinfo {author} {\bibfnamefont {D.}~\bibnamefont
  {Forster}}, \bibinfo {author} {\bibfnamefont {D.~R.}\ \bibnamefont
  {Nelson}},\ and\ \bibinfo {author} {\bibfnamefont {M.~J.}\ \bibnamefont
  {Stephen}},\ }\bibfield  {title} {\bibinfo {title} {Large-distance and
  long-time properties of a randomly stirred fluid},\ }\href@noop {} {\bibfield
   {journal} {\bibinfo  {journal} {Phys. Rev. A}\ }\textbf {\bibinfo {volume}
  {16}},\ \bibinfo {pages} {732} (\bibinfo {year} {1977})}\BibitemShut
  {NoStop}%
\bibitem [{\citenamefont {Huse}\ \emph {et~al.}(1985)\citenamefont {Huse},
  \citenamefont {Henley},\ and\ \citenamefont {Fisher}}]{Huse85}%
  \BibitemOpen
  \bibfield  {author} {\bibinfo {author} {\bibfnamefont {D.~A.}\ \bibnamefont
  {Huse}}, \bibinfo {author} {\bibfnamefont {C.~L.}\ \bibnamefont {Henley}},\
  and\ \bibinfo {author} {\bibfnamefont {D.~S.}\ \bibnamefont {Fisher}},\
  }\bibfield  {title} {\bibinfo {title} {Huse, Henley, and Fisher respond},\
  }\href@noop {} {\bibfield  {journal} {\bibinfo  {journal} {Phys. Rev. Lett.}\
  }\textbf {\bibinfo {volume} {55}},\ \bibinfo {pages} {2924} (\bibinfo {year}
  {1985})}\BibitemShut {NoStop}%
\bibitem [{\citenamefont {Krug}\ \emph {et~al.}(1992)\citenamefont {Krug},
  \citenamefont {Meakin},\ and\ \citenamefont {Halpin-Healy}}]{Krug92}%
  \BibitemOpen
  \bibfield  {author} {\bibinfo {author} {\bibfnamefont {J.}~\bibnamefont
  {Krug}}, \bibinfo {author} {\bibfnamefont {P.}~\bibnamefont {Meakin}},\ and\
  \bibinfo {author} {\bibfnamefont {T.}~\bibnamefont {Halpin-Healy}},\
  }\bibfield  {title} {\bibinfo {title} {Amplitude universality for driven
  interfaces and directed polymers in random media},\ }\href@noop {} {\bibfield
   {journal} {\bibinfo  {journal} {Phys. Rev. A}\ }\textbf {\bibinfo {volume}
  {45}},\ \bibinfo {pages} {638} (\bibinfo {year} {1992})}\BibitemShut
  {NoStop}%
\bibitem [{\citenamefont {Amar}\ and\ \citenamefont {Family}(1990)}]{Amar90}%
  \BibitemOpen
  \bibfield  {author} {\bibinfo {author} {\bibfnamefont {J.~G.}\ \bibnamefont
  {Amar}}\ and\ \bibinfo {author} {\bibfnamefont {F.}~\bibnamefont {Family}},\
  }\bibfield  {title} {\bibinfo {title} {Diffusion annihilation in one
  dimension and kinetics of the {I}sing model at zero temperature},\
  }\href@noop {} {\bibfield  {journal} {\bibinfo  {journal} {Phys. Rev. A}\
  }\textbf {\bibinfo {volume} {41}},\ \bibinfo {pages} {3258} (\bibinfo {year}
  {1990})}\BibitemShut {NoStop}%
\bibitem [{\citenamefont {Beccaria}\ and\ \citenamefont
  {Curci}(1994)}]{Beccaria94}%
  \BibitemOpen
  \bibfield  {author} {\bibinfo {author} {\bibfnamefont {M.}~\bibnamefont
  {Beccaria}}\ and\ \bibinfo {author} {\bibfnamefont {G.}~\bibnamefont
  {Curci}},\ }\bibfield  {title} {\bibinfo {title} {Numerical simulation of the
  {K}ardar--{P}arisi--{Z}hang equation},\ }\href@noop {} {\bibfield  {journal}
  {\bibinfo  {journal} {Phys. Rev. E}\ }\textbf {\bibinfo {volume} {50}},\
  \bibinfo {pages} {4560} (\bibinfo {year} {1994})}\BibitemShut {NoStop}%
\bibitem [{\citenamefont {Lam}\ and\ \citenamefont
  {Shin}(1998{\natexlab{a}})}]{Lam98a}%
  \BibitemOpen
  \bibfield  {author} {\bibinfo {author} {\bibfnamefont {C.-H.}\ \bibnamefont
  {Lam}}\ and\ \bibinfo {author} {\bibfnamefont {F.~G.}\ \bibnamefont {Shin}},\
  }\bibfield  {title} {\bibinfo {title} {Improved discretization of the
  {Kardar}--{Parisi}--{Zhang} equation},\ }\href@noop {} {\bibfield  {journal}
  {\bibinfo  {journal} {Phys. Rev. E}\ }\textbf {\bibinfo {volume} {58}},\
  \bibinfo {pages} {5592} (\bibinfo {year} {1998}{\natexlab{a}})}\BibitemShut
  {NoStop}%
\bibitem [{\citenamefont {Buceta}(2005)}]{Buceta05}%
  \BibitemOpen
  \bibfield  {author} {\bibinfo {author} {\bibfnamefont {R.~C.}\ \bibnamefont
  {Buceta}},\ }\bibfield  {title} {\bibinfo {title} {Generalized discretization
  of the {Kardar}--{Parisi}--{Zhang} equation},\ }\href@noop {} {\bibfield
  {journal} {\bibinfo  {journal} {Phys. Rev. E}\ }\textbf {\bibinfo {volume}
  {72}},\ \bibinfo {pages} {017701} (\bibinfo {year} {2005})}\BibitemShut
  {NoStop}%
\bibitem [{\citenamefont {Lam}\ and\ \citenamefont
  {Shin}(1998{\natexlab{b}})}]{Lam98b}%
  \BibitemOpen
  \bibfield  {author} {\bibinfo {author} {\bibfnamefont {C.-H.}\ \bibnamefont
  {Lam}}\ and\ \bibinfo {author} {\bibfnamefont {F.~G.}\ \bibnamefont {Shin}},\
  }\bibfield  {title} {\bibinfo {title} {Anomaly in numerical integrations of
  the {Kardar}--{Parisi}--{Zhang} equation},\ }\href@noop {} {\bibfield
  {journal} {\bibinfo  {journal} {Phys. Rev. E}\ }\textbf {\bibinfo {volume}
  {57}},\ \bibinfo {pages} {6506} (\bibinfo {year}
  {1998}{\natexlab{b}})}\BibitemShut {NoStop}%
\bibitem [{\citenamefont {Giada}\ \emph {et~al.}(2002)\citenamefont {Giada},
  \citenamefont {Giacometti},\ and\ \citenamefont {Rossi}}]{Giada02}%
  \BibitemOpen
  \bibfield  {author} {\bibinfo {author} {\bibfnamefont {L.}~\bibnamefont
  {Giada}}, \bibinfo {author} {\bibfnamefont {A.}~\bibnamefont {Giacometti}},\
  and\ \bibinfo {author} {\bibfnamefont {M.}~\bibnamefont {Rossi}},\ }\bibfield
   {title} {\bibinfo {title} {Pseudospectral method for the
  {K}ardar--{P}arisi--{Z}hang equation},\ }\href@noop {} {\bibfield  {journal}
  {\bibinfo  {journal} {Phys. Rev. E}\ }\textbf {\bibinfo {volume} {65}},\
  \bibinfo {pages} {036134} (\bibinfo {year} {2002})}\BibitemShut {NoStop}%
\bibitem [{\citenamefont {Gallego}\ \emph {et~al.}(2007)\citenamefont
  {Gallego}, \citenamefont {Castro},\ and\ \citenamefont
  {L\'opez}}]{Gallego07}%
  \BibitemOpen
  \bibfield  {author} {\bibinfo {author} {\bibfnamefont {R.}~\bibnamefont
  {Gallego}}, \bibinfo {author} {\bibfnamefont {M.}~\bibnamefont {Castro}},\
  and\ \bibinfo {author} {\bibfnamefont {J.~M.}\ \bibnamefont {L\'opez}},\
  }\bibfield  {title} {\bibinfo {title} {Pseudospectral versus
  finite-difference schemes in the numerical integration of stochastic models
  of surface growth},\ }\href@noop {} {\bibfield  {journal} {\bibinfo
  {journal} {Phys. Rev. E}\ }\textbf {\bibinfo {volume} {76}},\ \bibinfo
  {pages} {051121} (\bibinfo {year} {2007})}\BibitemShut {NoStop}%
\bibitem [{\citenamefont {Ascher}\ \emph {et~al.}(1995)\citenamefont {Ascher},
  \citenamefont {Ruuth},\ and\ \citenamefont {Wetton}}]{Ascher95}%
  \BibitemOpen
  \bibfield  {author} {\bibinfo {author} {\bibfnamefont {U.~M.}\ \bibnamefont
  {Ascher}}, \bibinfo {author} {\bibfnamefont {S.~J.}\ \bibnamefont {Ruuth}},\
  and\ \bibinfo {author} {\bibfnamefont {B.~T.~R.}\ \bibnamefont {Wetton}},\
  }\bibfield  {title} {\bibinfo {title} {Implicit-{Explicit} methods for
  time-dependent partial differential equations},\ }\href@noop {} {\bibfield
  {journal} {\bibinfo  {journal} {SIAM J. Numer. Anal.}\ }\textbf {\bibinfo
  {volume} {32}},\ \bibinfo {pages} {797} (\bibinfo {year} {1995})}\BibitemShut
  {NoStop}%
\bibitem [{\citenamefont {Canuto}\ \emph {et~al.}(1988)\citenamefont {Canuto},
  \citenamefont {Hussaini}, \citenamefont {Quarteroni},\ and\ \citenamefont
  {Jr.~Zang}}]{Canuto88}%
  \BibitemOpen
  \bibfield  {author} {\bibinfo {author} {\bibfnamefont {C.}~\bibnamefont
  {Canuto}}, \bibinfo {author} {\bibfnamefont {M.~Y.}\ \bibnamefont
  {Hussaini}}, \bibinfo {author} {\bibfnamefont {A.~M.}\ \bibnamefont
  {Quarteroni}},\ and\ \bibinfo {author} {\bibfnamefont {T.~A.}\ \bibnamefont
  {Zang~Jr.}},\ }\href@noop {} {\emph {\bibinfo {title} {Spectral Methods in
  Fluid Dynamics}}}\ (\bibinfo  {publisher} {Springer-{Verlag} {Berlin}
  {Heidelberg}},\ \bibinfo {year} {1988})\BibitemShut {NoStop}%
\bibitem [{\citenamefont {Spalart}\ \emph {et~al.}(1991)\citenamefont
  {Spalart}, \citenamefont {Moser},\ and\ \citenamefont {Rogers}}]{Spalart91}%
  \BibitemOpen
  \bibfield  {author} {\bibinfo {author} {\bibfnamefont {P.~R.}\ \bibnamefont
  {Spalart}}, \bibinfo {author} {\bibfnamefont {R.~D.}\ \bibnamefont {Moser}},\
  and\ \bibinfo {author} {\bibfnamefont {M.~M.}\ \bibnamefont {Rogers}},\
  }\bibfield  {title} {\bibinfo {title} {Spectral methods for the
  {Navier}-{Stokes} equations with one infinite and two periodic directions},\
  }\href@noop {} {\bibfield  {journal} {\bibinfo  {journal} {J. Comp. Phys.}\
  }\textbf {\bibinfo {volume} {96}},\ \bibinfo {pages} {297 } (\bibinfo {year}
  {1991})}\BibitemShut {NoStop}%
\bibitem [{\citenamefont {Horowitz}\ \emph {et~al.}(2001)\citenamefont
  {Horowitz}, \citenamefont {Monetti},\ and\ \citenamefont
  {Albano}}]{Horowitz01}%
  \BibitemOpen
  \bibfield  {author} {\bibinfo {author} {\bibfnamefont {C.~M.}\ \bibnamefont
  {Horowitz}}, \bibinfo {author} {\bibfnamefont {R.~A.}\ \bibnamefont
  {Monetti}},\ and\ \bibinfo {author} {\bibfnamefont {E.~V.}\ \bibnamefont
  {Albano}},\ }\bibfield  {title} {\bibinfo {title} {Competitive growth model
  involving random deposition and random deposition with surface relaxation},\
  }\href@noop {} {\bibfield  {journal} {\bibinfo  {journal} {Phys. Rev. E}\
  }\textbf {\bibinfo {volume} {63}},\ \bibinfo {pages} {066132} (\bibinfo
  {year} {2001})}\BibitemShut {NoStop}%
\bibitem [{\citenamefont {Kim}\ and\ \citenamefont {Kosterlitz}(1989)}]{Kim89}%
  \BibitemOpen
  \bibfield  {author} {\bibinfo {author} {\bibfnamefont {J.~M.}\ \bibnamefont
  {Kim}}\ and\ \bibinfo {author} {\bibfnamefont {J.~M.}\ \bibnamefont
  {Kosterlitz}},\ }\bibfield  {title} {\bibinfo {title} {Growth in a restricted
  solid-on-solid model},\ }\href@noop {} {\bibfield  {journal} {\bibinfo
  {journal} {Phys. Rev. Lett.}\ }\textbf {\bibinfo {volume} {62}},\ \bibinfo
  {pages} {2289} (\bibinfo {year} {1989})}\BibitemShut {NoStop}%
\end{thebibliography}
\end{document}